\documentclass{actasen}
\usepackage{amssymb}

\begin{document}

%%ÉèÖÃÊ×Ò³Ò³Âë
\setcounter{page}{1}

\Volume{2015}{39?}% Äê¡¢¾í

%%ҳüÉèÖÃ

\runheading{XU Ren-xin}%

\title{The many faces of strange matter:$^{\S}$\\
{\it compact stars, cosmic rays, and dark matter}}

\footnotetext{$^{\S}$ Supported by the 973 program (No.
2012CB821801), NNSFC (11225314), and XTP XDA04060604.

%Received 2013--05--02; revised version 2013--08--06

\hspace*{5mm}$^{\star}$ {\tt Email}: r.x.xu@pku.edu.cn\\

\noindent 0275-1062/01/\$-see front matter $\copyright$ 2011 Elsevier
Science B. V. All rights reserved. %%

\noindent PII: }

\enauthor{XU Ren-xin$^{\star}$ }{School of Physics and State Key Laboratory of Nuclear Physics and Technology, and\\
Kavli Institute for Astronomy and Astrophysics, Peking University, Beijing 100871, China}

\abstract{%
The state of cold bulk matter at around nuclear density depends on the fundamental strong interaction between quarks at low-energy scale, so-called non-perturbative quantum chromo-dynamics. Such kind of matter is conjectured to be condensed matter of 3-flavour ($u$, $d$ and $s$) quark clusters in this note, being manifested in the form of compact stars, cosmic rays, and even dark matter.
}%


\maketitle

\section{Introduction: Bigger is different!}

There are six flavours of quark in the well-tested standard model of particle physics, half ($u$, $d$ and $s$, with current masses smaller than $\sim 100$ MeV) are light and the other half ($c$, $t$ and $b$, masses $>1$ GeV) are heavy.
However, the common material of the world today is built only from two flavours ($u$ and $d$) of quark, rather than three ($u$, $d$ and $s$), although the zero energy $\sim \hbar c/$fm of quarks inside atomic nuclei should be higher than the mass difference ($\sim 100$ MeV) between $s$-quark and $u$/$d$-quarks. Why is our world 2-flavour symmetric?

The answer could be: micro-nuclei are too small to have 3-flavour symmetry, but {\em bigger is different}!
In fact, rational thinking about stable strangeness dates back to 1970s,\rf{1} in which Bodmer speculated that so-called ``collapsed nuclei'' especially with strangeness could be energetically favored if baryon number $A>A_{\min}$, but without quantitative estimation of the minimal number, $A_{\min}$.
Bulk matter composed of almost {\em free} quarks ($u$, $d$, and $s$) was focused on,\rf{2,3} even for astrophysical manifestations.\rf{4,5}
One essential point, however, is whether the color coupling between quarks is still perturbative in cold dense matter at a few nuclear densities although the asymptotic freedom is well recognized.
The strong force there might render quarks grouped in so-called {\em quark-clusters}, forming a nucleus-like strange object\rf{6} with 3-flavour symmetry if it is big enough that relativistic electrons are inside (i.e., $A>A_{\rm min}\simeq 10^9$).\footnote{
$^1$%
Strange matter formation ({\em strangenization}) could be considered as an extension of neutronization of $e+p\longleftrightarrow n+\nu_e$, in both of which the weak interaction plays an important role by electrons' participation if the scale $>\lambda_{\rm c}=h/m_{\rm e}c=0.024$ \AA. Strangenization has the advantages of (1) minimizing the electron's contribution of kinetic energy and (2) maximizing the flavour number, from 2 to 3. Bigger is then different.
} %
Anyway, we could simply call 3-flavour baryonic matter as {\em strange matter}, in which the constituent quarks could be either itinerant or localized.

Where could strange matter exist? How does it manifest? Can one discover it? These are questions we are trying to answer here.
Strange matter could be produced after cosmic hadronization (strange
nugget) or during a collapse event where normal baryonic matter is
intensely compressed by gravity (general core-collapse $\rightarrow$
strange star, degenerate O-Ne-Mg matter collapse $\rightarrow$ low-mass
strange star even strange planet).
Strange stars manifesting themselves as pulsar-like objects are self-bound on surface by strong force, which distinguish themselves significantly from self-gravitating neutron stars
(e.g., the sharp difference in mass-radius relation).
In this short note, we are focusing on strange quark-cluster matter
by which we could understand realistic dense matter better
than by strange quark matter.

\section{The mass spectrum of strange matter}

There are dark energy, dark matter and baryonic matter in the Universe, and it is usually known that the baryonic part is now in the form of atoms.
The conjectured {\em strange matter} could be simply condense matter of quark-clusters with strangeness\rf{6} and may manifest itself as a variety of objects with a broad mass spectrum, including compact objects, cosmic rays and even dark matter.

There are probably two or three channels to generate strange matter.
(1) The order of the QCD phase transition, during which quarks and gluons condense to form a gas of hadrons in the early Universe ($\sim 10~\mu$s), is still a matter of debate. Nevertheless, the transition could be of first order if quark clustering occurs, and strange nuggets may thus survive even after evaporation and boiling (suppressed by clustering). The baryon number of conjectured relic nuggets could be $ 10^{\sim (9-49)}$. Strange nuggets as cold quark matter may favour the formation of seed black holes in primordial halos,\rf{7} alleviating the current difficulty of quasars at redshift as high as $z \sim 6$.
(2) Strange matter could be the rump left behind after a collapse event where normal baryonic matter is intensely compressed by gravity. Besides general core-collapse of massive evolved star,
either O-Ne-Mg (stable $\alpha$-nuclei) degenerate core of evolved medium star or O-Ne-Mg white dwarf in binary (induced by accretion) could also collapse if neutronization (and further strangenization), rather than nuclear fusion, dominates. Strange nuggets might be ejected from the turbulent surface of such a nascent strange compact remnant.
(3) A massive strange star or a black hole could form via merging binary strange stars, during which strange nuggets/strangelets would be produced as cosmic rays.\rf{8}
No r-process nucleosynthesis occurs during the coalescence.
A mass spectrum of strange matter is speculated in Fig.~1.
\begin{figure}[tbph]
\centering
{\includegraphics[angle=0,width=13cm]{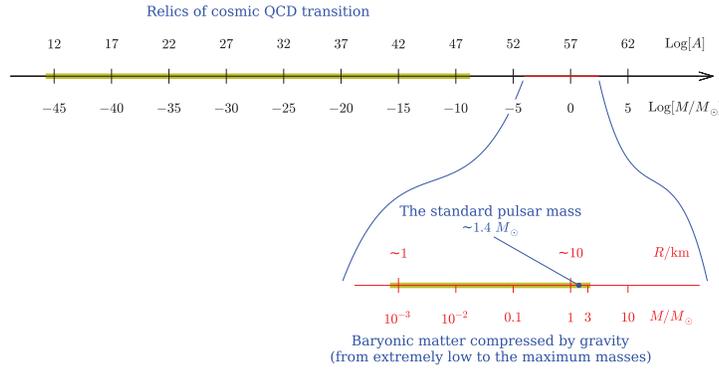}}
%\vspace{-2mm}
\caption{A speculated mass spectrum of strange objects (strange star, planet, and nugget/strangelet).}
    \end{figure}

As a first step, clear evidence for strange matter could be obtained by studying pulsar-like compact star, but it is worth noting that strange quark-cluster star and strange quark star behave very differently.
Besides having a stiffer equation of state, a strange quark-cluster star has a strangeness barrier on the surface, that would be essential for understanding the observations of X-ray bursters and even the symbiotic X-ray system 4U 1700+24.\rf{9}

\section{Conclusions}

Strange matter is conjectured to be condensed matter of 3-flavour quark-clusters, which could manifest itself as compact stars, cosmic rays, and even dark matter.
Future advanced facilities (e.g., FAST) would provide opportunity to find solid evidence for strange matter.


\begin{thebibliography}{999}
\bibitem{1} Bodmer A R. Phys. Rev., 1971, D4, 16

\bibitem{2} Itoh N. Prog. Theor. Phys., 1970, 44, 291

\bibitem{3} Witten E. Phys. Rev., 1984, D30, 272

\bibitem{4} Alcock C, Farhi E, Olinto A. 1986, ApJ, 310, 261

\bibitem{5} Haensel P, Zdunik J L, Schaeffer R. 1986, A\&A, 160, 121

\bibitem{6} Xu R X. ApJ, 2003, 596, L59

\bibitem{7} Lai X Y, Xu R X. JCAP, 2010, 5, 28

\bibitem{8} Madsen J. Phys. Rev., 2005, D71, 014026

\bibitem{9} Xu R X. RAA, 2014, 14, 617

\end{thebibliography}
\end{document}